# Shocks in Supersonic Sand

Erin C. Rericha,[1] Chris Bizon,[2] Mark D. Shattuck,[1] and Harry L. Swinney[1]

[1] Center for Nonlinear Dynamics, The University of Texas at Austin, Austin, TX 78712
[2]Colorado Research Associates Division, Northwest Research Assoc, Boulder, CO 80301

We measure time-averaged velocity, density, and temperature fields for steady granular flow past a wedge. We find the flow to be supersonic with a speed of granular pressure disturbances (sound speed) equal to about 10% of the flow speed, and we observe shocks nearly identical to those in a supersonic gas. Molecular dynamics simulations of Newton's laws yield fields in quantitative agreement with experiment. A numerical solution of Navier-Stokes-like equations agrees with a molecular dynamics simulation for experimental conditions excluding wall friction.

Shocks form around an object such as a bullet or an aircraft when the speed of the object relative to the incident flow exceeds the speed of sound in the fluid [1] Shocks analogous to those that form in fluid flows also occur in flows of macroscopic particles such as sand grains [2]. The usual theoretical approach to understanding granular flows is dense gas kinetic theory, treating the constituent grains as colliding, inelastic hard spheres. As in standard dense gas kinetic theory, flows of particles that are described by Newton's laws are modeled with a Boltzmann equation, which in turn leads to Navier-Stokes-like continuum equations. For granular media these continuum equations contain a term that describes the overall energy loss due to inelastic collisions [3,4].

The inelastic collisions in a granular flow reduce the relative velocities of the grains; consequently the local granular temperature, defined as the variance of the local velocity distribution, decreases [5]. Whether the fluid is composed of grains or molecules, the speed of sound depends on the speed of the component particles and therefore on the temperature. (For a granular fluid, the speed of sound in the interstitial air is irrelevant; a granular fluid has the same sound speed even in a vacuum.) Since inelastic collisions dissipate temperature, the speed of sound in a granular flow decreases. In the absence of further heating, a granular flow becomes supersonic as it progresses, i.e., the mean particle velocity surpasses the speed of sound [6]. Thus, shocks form in granular systems for common rather than extreme conditions whenever the flow encounters an obstacle.

Experimental tests of the continuum equations for granular media have been limited to quantities such as the particle diffusion coefficient [7], stress-strain curves [8], and a localized velocity profile [5,9]. Further, continuum theories [3,4] assume an equilibrium Maxwell-Boltzmann velocity distribution, no velocity correlations (molecular chaos), small dissipation, and a clear scale separation between the microscopic and macroscopic. Recent molecular dynamics simulations [10], analyses [11], and experiments [12] question the domain over which these assumptions are justified. As new versions of kinetic theory emerge to deal with these issues [13], detailed experimental tests are required to sort out under what conditions each assumption is justified. We present here a quantitative study of shocks and expansion fans in a granular flow. We first present the experimental method and observations, and then we present the method and results of molecular dynamics and continuum equations simulations.

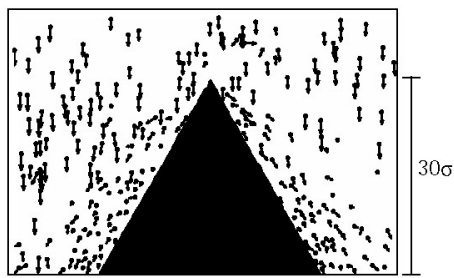

FIG 1. An observed image of granular flow incident downward on a wedge, where the particle positions and velocities (denoted by arrows) are determined from images separated by 1 ms. The longest arrow corresponds to a velocity of 1.65 m/s. The figure, 68$\sigma$ by 45$\sigma$ (where $\sigma$ is the particle diameter), shows the top 30$\sigma$ of a wedge of total height 100 $\sigma$.

*Experiment.* Stainless steel particles of diameter $\sigma$ (1.2 mm) fall under gravity past a wedge sandwiched between glass plates separated by 1.6$\sigma$ (Fig. 1). Particles are initially uniformly distributed on a conveyor belt 400$\sigma$ wide. The particles fall off the conveyor belt into a hopper and enter the cell a distance of 42$\sigma$ above the wedge tip with an initial volume fraction of 0.018. From high-speed digital images particle positions are determined to 0.023 mm (0.019$\sigma$) and velocities to 0.023 m/s (2\% of typical particle velocities). Velocities and positions are time-averaged over 16000 frames to get the bulk flow fields of velocity, area fraction, and granular kinetic temperature. The typical sound speed, determined using our measured fields in the continuum theory expression [5,14], is 0.09 m/s. The flow enters the cell with a Mach number (ratio of the flow speed to the sound speed) of 7 and reaches a Mach number of 12 at the tip of the wedge.

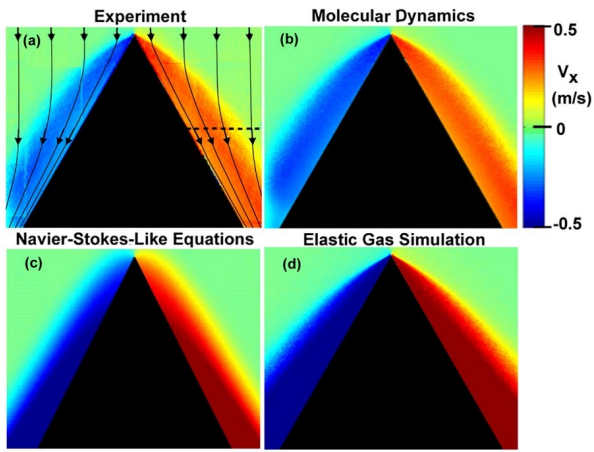

FIG 2. Horizontal component of the velocity field of a granular flow incident downward on a wedge, determined by three methods: (a) experiment, (b) molecular dynamics simulation of Newton's Laws, and (c) integration of Navier-Stokes-like equations. A molecular dynamics simulation of a hard sphere elastic gas (d) is included for a reference. Each picture shows a region $130\sigma$ by $104\sigma$. The solid lines with arrows denote streamlines. Quantitative comparisons along the dashed line in (a) are shown in Fig. 4 and 5.

*Observations*. As in supersonic gases, a granular flow forms a shock when it encounters an obstacle (Fig. 1). The no-flux boundary condition at the wedge surface requires the flow to change direction to pass around the wedge. Since the flow is supersonic, streamlines must alter rapidly, within a particle mean free path. The shock separates flow that is unaware of the obstacle and consequently has no horizontal velocity, from flow that is aware of the obstacle and has acquired a horizontal velocity component (Figs. 1 and 2). After the shock the flow has a higher volume fraction, higher temperature, and lower mean velocity, just as for shocks in gas flows. The shock does not require inelasticity to form as can be seen by comparing the supersonic inelastic and elastic gases in Fig. 2(b) and Fig. 2(d) respectively. Because of gravity, both the inelastic and elastic gases do not extend out at a constant angle but curve toward the wedge.

At the bottom of the wedge we observe an expansion fan, as illustrated in Fig. 3. An expansion fan, which is the opposite of a shock, forms when the volume available to a supersonic flow suddenly increases rather than decreases; this phenomenon is well studied in gas flows [1]. In a fan the density and temperature decrease and the Mach number increases. In contrast to a shock, an expansion fan is not a rapid change; instead, the fan radiates from a point, which in the present case is the bottom corner of the wedge.

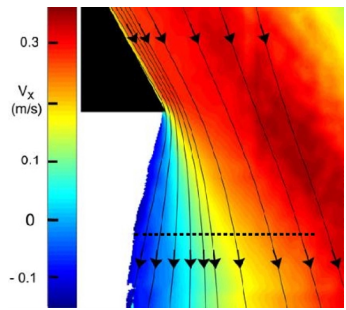

FIG 3. The horizontal velocity field measured for the expansion fan formed when the supersonic granular flow reaches the bottom of the wedge. The solid lines indicate selected streamlines. The total height of the region shown is $55\sigma$. The white region below the wedge has too few particles to determine the velocity. The horizontal velocity profile along the dashed line is shown in Fig.4(c).

*Theory.* The simple geometry and steady state behavior of the present experiment provide an ideal test of theoretical descriptions of granular flow. We compare the experimental observations to two levels of granular flow theory: the microscopic dynamics of the particles governed by Newton's laws are solved using an event-driven molecular dynamics method (Fig. 2(b)), and Navier-Stokes-like continuum equations for the macroscopic granular fields are numerically solved by a finite difference method (Fig. 2(c)). In both simulations the coefficient of restitution $e$ is 0.97. We now describe each method in turn.

A three-dimensional, event-driven molecular dynamics simulation computes the motion of each particle [15]. Between collisions, particles follow parabolic motion; when two particles collide, they undergo an instantaneous binary collision that conserves momentum and angular momentum but dissipates kinetic energy. The collision is treated using the operator due to Walton [16], which characterizes the collision in terms of coefficients of restitution $e$ and friction $\mu$. We use the same value of $\mu$ (0.15) to model inter-particle and particle-wall collisions. The ratio of temperature perpendicular to the wall to that parallel to the wall, $\alpha$, is set to 0.8. These parameters, which are not measured in the experiment, are set to provide good agreement with the experiment throughout the full plane, including the incident free-stream velocity. Changing the values of these parameters changes the magnitude of the flow behind the shock but does not affect the qualitative behavior of the flow. Reducing the coefficient of restitution from 0.97 to 0.90 reduces the maximum horizontal velocity by 27%.

The continuum equations proposed by Jenkins and Richman differ from Navier-Stokes equations for an ordinary non-isothermal, compressible, dense fluid only by a modified equation of state and by the addition of a temperature loss rate term arising from the dissipative collisions [3]. The equations are integrated using a second order accurate finite difference scheme on a two-dimensional rectangular grid, assuming the flow in the third direction is uniform. Boundary conditions at the inlet are taken from the experiment, and at the outlet are free. On the wedge, slip velocity conditions are used in which the ratio of the tangential to normal strain rate is set to 1. The heat flux on the wedge boundary is proportional to the density and the temperature raised to the 3/2 power [17]. The proportionality constant is set to 0.2 by measuring the average heat flux from the molecular dynamics simulations. Euler time stepping is used to increment time until a steady state is reached in which the horizontally averaged mass flux is constant to 0.01%.

9.8 m/s². This difference could be due to the air inside the thin channel, wall friction, or a combination of the two effects. The molecular dynamics simulation is three-dimensional, including the confining walls, and allows for friction during ball-wall collisions. The computational time for a three-dimensional continuum simulation is prohibitive, and a two-dimensional model for frictional drag does not exist. Because of this effect, we compare the experiment first to the molecular dynamics simulation with wall friction, and then we compare the molecular dynamics and continuum equations simulations calculated without wall friction.

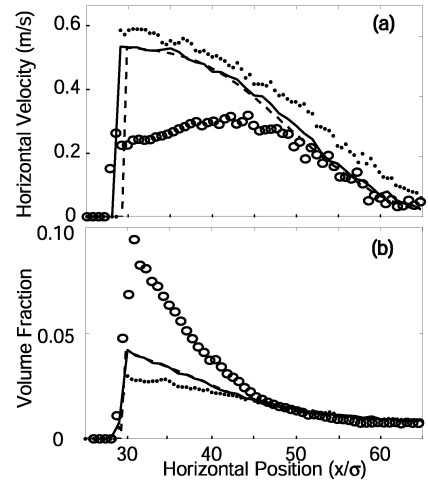

FIG 5. Comparison of shock profiles for granular flow past a wedge obtained from molecular dynamics (solid line) and the continuum equations (dashed line) assuming no friction along the confining sidewalls. A molecular dynamics simulation of elastic particles (dotted line) is included for comparison. (a) Horizontal velocity profile and (b) volume fraction along the dashed line in Fig. 2(a). Experimental measurements (circles) show the similar qualitative behavior but disagree quantitatively. The difference between the continuum simulation and the experiment is due to wall friction, but wall friction is not integral to shock formation itself.

*Comparison of experiment and theory.* In each approach we determine the time-averaged velocity, temperature, volume fraction, and Mach number fields. Results from the molecular dynamics simulation are compared with experiment in Fig. 4 for the horizontal velocity component and volume fraction. The root mean square difference between experiment and simulations is less than 2\% for volume fraction, velocity, and temperature throughout the flow. Our molecular dynamics simulation agrees well with experiment; however, it does not provide the intuition gained from a Navier-Stokes-like description of the macroscopic flow. Our continuum simulation is two-dimensional, unlike the

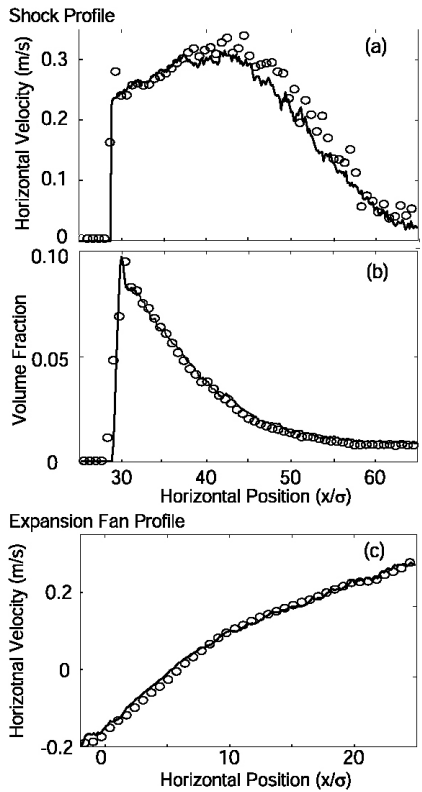

FIG 4. Shock and expansion fan profiles for granular flow past a wedge measured in the experiment (circles) are compared with the predictions from molecular dynamics (solid line): (a) Horizontal component of velocity and (b) volume fraction along the dashed line in Fig. 2(a). (c) Horizontal component of velocity along the dashed line in Fig. 3.

The confining glass side walls in the experiment affect the flow; the measured average acceleration of the free stream inside the cell is 8.9 m/s², while a particle outside the cell falls with the expected downward acceleration of molecular dynamics simulation. Therefore, the continuum simulations do not capture the interactions between the granular fluid and the confining glass plates. We compare the continuum simulations to the experiment indirectly using the molecular dynamics simulation in which the wall drag is also neglected, as shown in Fig. 5. The two simulations, one based on individual particles and the other based on a continuous medium, agree to within 1\% in the bulk with a maximum error of 10\% in a region a few σ from the tip of the wedge. The larger errors indicate a problem with the simple boundary conditions used in this work. However, the excellent agreement in the bulk confirms the applicability of continuum equations and validates the kinetic theory approach used to derive them. Since the molecular dynamics simulations with wall friction agree with the experiment, and the continuum simulations agree with molecular dynamics simulations without wall friction, we attribute the difference between the continuum model and the experiment to wall friction.

In conclusion, our experiments on granular flow past a wedge reveal shocks that are analogous to those in gas flows. We find that both molecular dynamics and continuum equations predict the quantitative behavior of a supersonic granular flow past an obstacle in the regime of low dissipation and low volume fraction. The disagreement between experiment and continuum modeling is, as we show in Fig. 5, a consequence of the confining sidewalls. While the thin cell geometry is useful in experiments because it facilitates imaging, most real applications will be fully three dimensional with negligible wall interactions; in such cases, our results indicate that continuum equations should provide an excellent model. Further work should examine flows with higher dissipation and volume fraction, where failure of the assumptions of theory is expected to be pronounced. In that regime, Monte Carlo simulations of the inelastic Boltzmann equation should provide physical insight into the flow [18]. Further, future experiments and analyses should examine the role of wall friction and air friction, simulations of the continuum equations should be extended to three dimensions, and better boundary conditions should be developed at the wedge tip.

We thank J. Burgess, J. de Bruyn, D. Goldman, J. Jenkins, B. Lewis, W. D. McCormick, S. J. Moon, and P. Umbanhowar for helpful suggestions. This research was supported by the Engineering Research Program of the Office of Basic Energy Sciences of the U.S. Department of Energy. CB was supported by NorthWest Research Associates.